# Tumorigenesis as a trauma response: the fragmentation of morphogenetic memory drives neoplastic dissociation

Author: Jordan Strasser[1*]

[1]Allen Discovery Center at Tufts University, Medford, MA 02155, USA

*Correspondence: jordanwstrasser@gmail.com

**Abstract**
The mitigation of stress is a key challenge for all biological systems. Conditions of unresolvable stress have been associated with a diverse array of pathologies, from cancer to post-traumatic stress disorder (PTSD). Here, I unify insights from evolutionary and developmental biology with trauma psychology to present a novel framework for tumorigenesis which synthesizes stress-perception, tissue dysfunction, and the hallmarks of neoplastic growth. This view carries therapeutic implications, suggesting a reintegrative approach that seeks to return cancer cells to the homeostatic control of the surrounding tissue.

**Introduction**

Over the past few decades, the conventional understanding of cancer has evolved from the gene-centric view of tumorigenesis to a more nuanced framework which recognizes that oncogenic mutations, in combination with permissive microenvironmental conditions, are required to unleash cellular proliferation and neoplastic growth (Bissell & Hines, 2011; Fidler, 2003; Hanahan & Weinberg, 2011). However, this modern understanding still does not fully account for the central role that *stress* plays in tumor initiation and evolution. Here, I argue that tumorigenesis may be best understood as a morphogenetic trauma-response: a stress-induced dissociative loop which reinforces anatomically intrusive behavior. To establish this claim, I first turn to evolutionary and developmental biology to show how the sharing of stress gives rise to the shared regulatory architecture of multicellularity. I then explore how the phenotypic 'hallmarks of cancer' naturally emerge from this architecture under conditions of unresolvable error (Hanahan & Weinberg, 2000). Finally, I draw on trauma psychology to demonstrate how tumor growth phenotypically mirrors the fragmented memory dynamics of post-traumatic stress disorder (PTSD). With this framework in place, I propose a new therapeutic paradigm for oncology rooted in the reprocessing of morphogenetic memory.

## 1. Morphogenesis: The Sharing of Stress Drives Multicellular Integration

<u>1.1 Building a cell</u>
For much of its history, life on earth consisted of single-celled organisms which each competed over local resources to maintain homeostasis—a narrow band of internal biochemical parameters which enabled their continued survival and reproduction (Fig. 1A) (Davies, 2016). If we broadly define *stress* as the discrepancy between a cell's homeostatic setpoints and its internal sensory states, then natural selection should favor cells that minimize this stress efficiently (Shreesha & Levin, 2024). Because effective stress reduction requires the ability to sense external conditions, this evolutionary pressure led to the development of signal transduction pathways—molecular circuits that converted environmental signals into intracellular states the cell could act upon. This marked the emergence of a basal form of *interpretation*, allowing the cell to transform sensory input into adaptive, goal-directed behavior (Lyon, 2006).

Over time, cells evolved increasingly sophisticated interpretive structures to accelerate stress minimization by linking specific sensory states to corresponding biochemical action programs. These structures are composed of protein-protein interactions, post-transcriptional and translational modifications, and epigenetic regulation (Deribe et al., 2010; Duncan et al., 2014; Nachtergaele & He, 2017; Rajagopala et al., 2014). Through feedback-driven plasticity, cells modulate these systems to construct a regulatory architecture—a learned *map* of stress-response relationships that enables homeostatic stability across diverse conditions (Applewhite et al., 1969; Lyon, 2015). Evidence for this internal map appears in a wide range of adaptive behaviors: cells adjust their motility (Adler & Tso, 1974; Webre et al., 2003), rewire their metabolic pathways (Goo et al., 2012; Mitchell et al., 2009), and modulate their gene expression in response to environmental change (Dhar et al., 2013; Hsieh & Wanner, 2010). As a temporally persistent information structure that guides environmentally appropriate behavior, this regulatory architecture forms the basis of cellular memory (Lyon, 2015).

1.2 Building a tissue

The emergence of multicellularity during the Metazoan Revolution has been linked to the evolution of mechanisms for sharing internal stress signals between cells (Shreesha & Levin, 2024). With the advent of paracrine signaling, cytoplasmic stress signals began to propagate between neighboring cells, functionally coupling individual homeostatic circuits by synchronizing error states (Plattner, 2017; Shreesha & Levin, 2024). As a result, individual interpretive architectures adapt through feedback to minimize stress not only within the cell, but across neighboring cells as well (Koseska & Bastiaens, 2017). By implicitly encoding each other's information-processing 'perspective,' these distributed networks construct a shared regulatory architecture which aligns local stress responses with emergent collective goals (Fig. 1B).

The molecular evolution of intercellular linkages—such as connexins, cadherins, and integrins—enabled cellular collectives to accelerate the learning rate of shared regulatory architectures by compressing the spatiotemporal distance between action and feedback (Fig. 1C). Connexins form gap junctions between cells, permitting the direct passage of small molecules and ions that mediate shared resting membrane voltages ($V_{mem}$) and internal cytoplasmic states (Mathews & Levin, 2017). Cadherins facilitate intercellular adhesion, enabling the direct transmission of epigenetically instructive forces across the actin cytoskeleton (Borghi et al., 2012; Hulpiau & Van Roy, 2009). Integrins anchor cells to a shared extracellular matrix (ECM), directly transmitting mechanical and biochemical distortions that influence nuclear architecture and epigenetic regulation (Nelson & Bissell, 2006; Zaidel-Bar, 2009). Together, these molecular mechanisms directly couple internal states across cells, giving rise to self-organizing tissues—physically integrated entities capable of responding to signals as a unified whole.

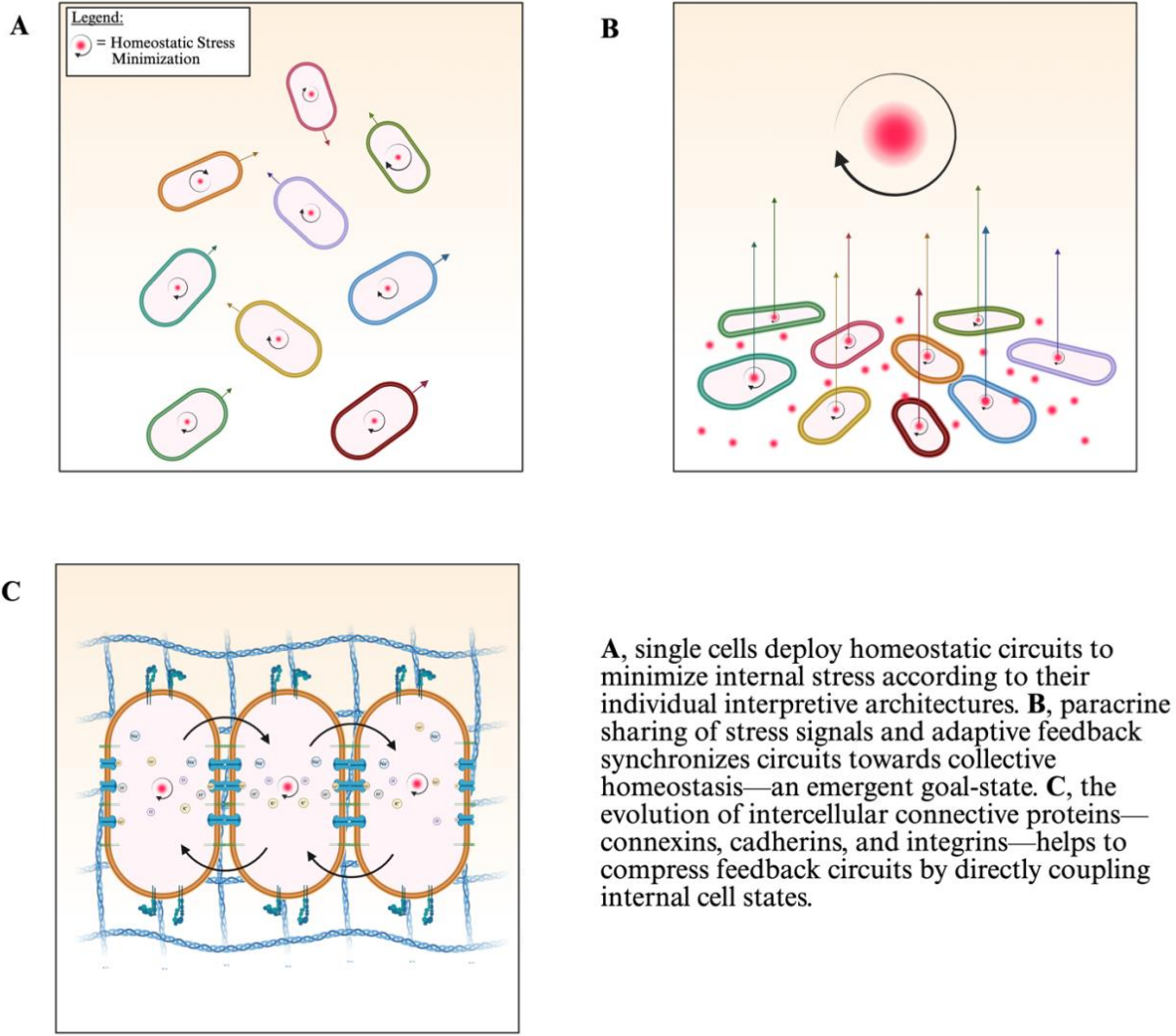

**Fig. 1: The Rise of Multicellularity**

**A**, single cells deploy homeostatic circuits to minimize internal stress according to their individual interpretive architectures. **B**, paracrine sharing of stress signals and adaptive feedback synchronizes circuits towards collective homeostasis—an emergent goal-state. **C**, the evolution of intercellular connective proteins—connexins, cadherins, and integrins—helps to compress feedback circuits by directly coupling internal cell states.

1.3 Building an organ

However, the immediate feedback provided by internal coupling can limit the system's ability to explore the space of stress-minimizing morphologies. Thus, cells evolved the capacity to modulate the weighting and connectivity of intercellular signaling structures through both epigenetic and post-translational gating mechanisms (Geng et al., 2012; Hotchin & Watt, 1992; Mathews & Levin, 2017). By spatially regulating the transmission of stress signals (Riol et al., 2021), tissues could specialize in discrete actions—such as secretion, absorption, and contraction—while remaining coordinated toward shared morphological endpoints (Ben-Moshe & Itzkovitz, 2019). Once these intercellular connections stabilize, they establish ***morphogenetic memory***: a temporally persistent topology of intercellular connectivity that guides tissue behavior across both developmental and regenerative timescales (Levin, 2012).

This memory is enacted through several interwoven mechanisms. Bioelectrical regionalization of resting membrane potential ($V_{mem}$) patterns large-scale morphological change in Metazoan development, from left-right symmetry (Aw et al., 2010; Pai et al., 2017) to organ-specific differentiation (Lobikin et al., 2015; Pai et al., 2012; Pai & Levin, 2022; Perathoner et al., 2014; Vandenberg et al., 2011) and plays a central role in regeneration (Adams et al., 2007; Levin, 2014; Levin et al., 2017; Pai et al., 2016). In parallel, cadherin- and integrin-based adhesion complexes generate spatially patterned mechanical forces across tissues, constraining gene expression and directing cell fate decisions during organogenesis (Ingber & Jamieson, 1985; Mammoto & Ingber, 2010; Murugan et al., 2021; Nelson & Bissell, 2006; Schmeichel et al., 1998). Morphogen gradients—such as those formed by Wnt, BMP and HOX proteins—overlay these networks to encode an anatomical coordinate system which orchestrates the spatial interpretation of developmental cues (Ashe & Briscoe, 2006; Chang et al., 2002; Gurdon et al., 1999; Niehrs, 2010; Rinn et al., 2006).

Morphogenetic memory serves as a shared reference frame through which developmental signals are interpreted to guide tissues toward stress-minimizing forms. Two key signals are growth factors (GFs) and damage-associated molecular patterns (DAMPS). By their relative concentrations and receptor-binding dynamics, these molecules act as *morphological error cues*—indicating discrepancies between the current tissue state and its target form (Aoki et al., 2017; Koseska & Bastiaens, 2020; Vénéreau et al., 2015). In response, cell collectives coordinate proliferation and differentiation until these signals are reduced to a morphostatic baseline that protects tissue integrity from damage, aging, and disease (Tarin, 1972).

However, morphogenetic memory is not fixed—it is plastic, and modifications to its architecture can fundamentally alter how error signals are interpreted by cellular collectives. In zebrafish, RNAi-mediated disruption of β-catenin gradients along the anterior-posterior axis creates a "missing tissue context" in which superficial wound signals are interpreted as cues to regenerate an entire ectopic limb (Owlarn et al., 2017). In planaria, pharmacological inhibition of gap junctions modifies bioelectrical connectivity, such that upon dissection, the tissue interprets the altered polarity as a cue to regenerate two heads (Oviedo et al., 2010). In Xenopus embryos, the induction of a specific range of hyperpolarized $V_{mem}$ values is sufficient to instruct surrounding cells to build an eye—even when the cells are located in gut, tail or lateral plate mesoderm regions (Pai et al., 2012). These cases demonstrate the latent plasticity of the tissue's interpretive frame: when morphogenetic memory is rewritten, the same error signals can yield dramatically different anatomical outcomes.

1.4 Building a body

As organisms evolved multi-organ systems with interconnected vasculature, local error correction via paracrine stress-sharing was expanded into body-wide coordination through endocrine signaling. Within this framework, specialized *physiological error signals*—such as catecholamines, glucocorticoids, reproductive hormones, and metabolic hormones—are interpreted by cell collectives to initiate local stress-minimizing actions that contribute to shared physiological goals across multiple organs. For instance, during acute threats to bodily integrity, catecholamines and glucocorticoids are interpreted by the heart, lungs, liver, muscle and eyes to trigger coordinated actions that mobilize the entire organism for fight-or-flight (Dickerson & Kemeny, 2004; Wortsman, 2002). Reproductive hormones are interpreted by distributed tissues as developmental error cues, signaling discrepancies between current tissue states and temporally programmed transitions across menstrual, estrous, or life-stage cycles (Pauerstein et al., 1978). Metabolic hormones—including insulin, thyroid hormones (T3/T4), leptin, and ghrelin—are decoded by diverse tissues to coordinate blood glucose, lipid storage, mitochondrial function and energy expenditure in support of whole-body homeostasis (Joshi et al., 2007; Klok et al., 2007; van der Spek et al., 2017). Though the sharing of stress enables organs to act in concert, prolonged exposure to unresolvable error can alter the very architectures through which tissues interpret physiological signals.

## 2. Tumorigenesis: Unresolvable Stress Drives Neoplastic Dissociation

<u>2.1 Breaking an organ</u>

Under normal physiological conditions, integrated tissues minimize stress across a wide range of variables while preserving their original structures. The heart can beat faster, renal tubules can increase secretion, and intestinal villi can reduce absorption—all without altering their underlying morphological properties (Fig. 2A). Although these anatomical forms appear stable, their underlying cell collectives retain the ability to adjust the strength and configuration of intercellular connections, dynamically modulating how tightly actions are coupled to feedback. In other words, tissues retain the ability to reshape the memory architecture which gives rise to morphological decision-making.

Chronic, unresolvable stress can cause tissues to weaken their intercellular connections and explore stress-minimizing morphologies beyond typical anatomical limits (Fig. 2B) (Levin, 2019). This adaptive search may yield metaplastic, hyperplastic, or dysplastic precancerous states—localized error-correction programs that have become uncoupled from whole-organ homeostasis (Faupel-Badger et al., 2024).

In Barret's esophagus, unmitigable stress in the form of acid reflux remodels the ECM and stroma of the squamous lining, prompting a metaplastic conversion to columnar cell types better suited to the acidic microenvironment (Strasser et al., 2025). While this adaptation temporarily mitigates stress for the epithelial lining, it markedly increases the risk of adenocarcinoma by up to two orders of magnitude (Spechler & Souza, 2014). In ductal carcinoma in situ (DCIS), persistent inflammation stiffens periductal collagen, disrupts myoepithelial adhesion, reprograms cancer-associated fibroblasts, and degrades the basement membrane, progressively distorting epithelial-stromal cross-talk (Acerbi et al., 2015; Casasent et al., 2022; Hu et al., 2008). As they gradually dissociate from the ductal-lobular system, luminal cells become hypersensitized to epidermal growth factors via HER2 upregulation, driving hyperplastic proliferation (Acerbi et al., 2015). And in inflammatory bowel disease, in response to unresolvable immune assault, gut epithelia will begin to alter morphology, positioning, and polarity, fostering dysplastic transformation and significantly increasing the risk of colorectal cancer (Zisman & Rubin, 2008).

These examples suggest that metaplasia, hyperplasia, and dysplasia are specific adaptations which emerge as the shared regulatory architecture between the affected tissue and parent organ gradually decouples. They also suggest that the tissue which begins to dissociate is the one most vulnerable to a particular dimension of unresolvable stress—whether chemical, mechanical, or immunological. Similar adaptations have been documented in a wide array of tissues, including the colon (Dornblaser et al., 2024), pancreas (Scheiman et al., 2015), prostate (Gurel et al., 2014), oral cavity (William Jr et al., 2023) and bone marrow (Yehudai-Resheff et al., 2019).

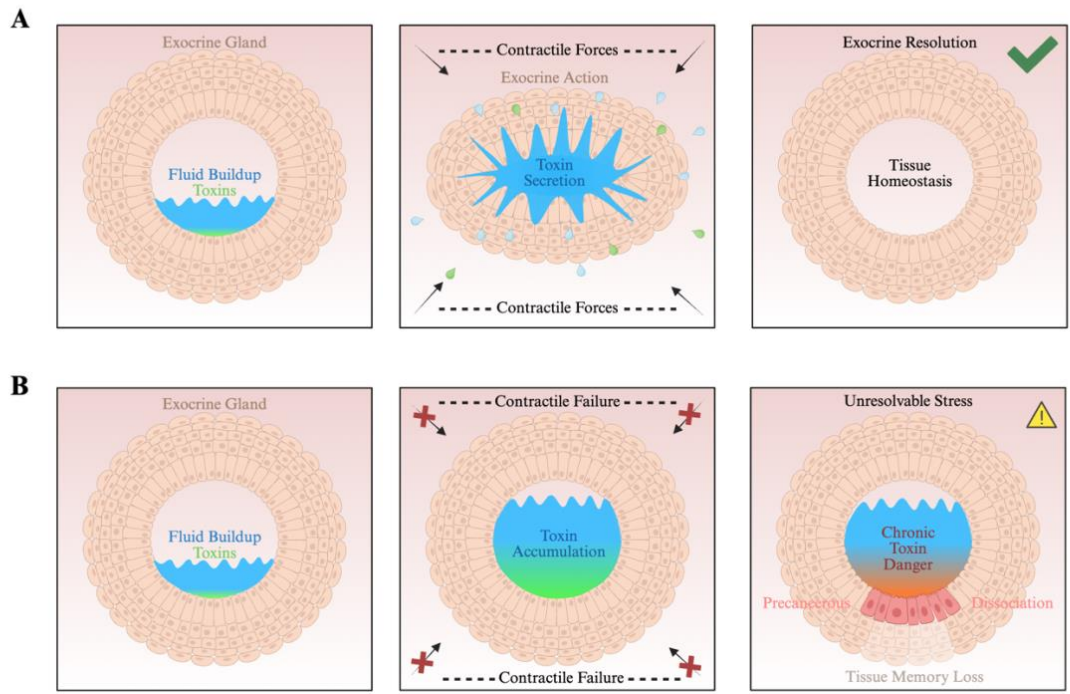

**Fig. 2: Unresolvable Tissue Stress Produces Precancerous Dissociation.**
**A, Collective Resolution.** Generic exocrine gland exerts contractile forces to effectively secrete toxins out of interior, leading to stress resolution and restored tissue homeostasis.
**B, Collective Failure.** Due to genetic mutation or structural limitation, contraction is prevented, leading to local toxin accumulation. Under chronically unresolvable stress, vulnerable tissue segments may begin to functionally decouple from whole-organ maintenance, eroding the integrity of local morphogenetic memory systems.

2.2 Breaking a tissue

If chronic, unresolvable stress persists, the tissue can cross a threshold into neoplasia—a state defined by the complete fragmentation of the body's shared regulatory architecture. Whether initiated by oncogenic mutations or microenvironmental constraints, prolonged error accumulation weakens intercellular coupling, eroding the feedback loops that align individual cell action with tissue-wide morphogenetic programs (Aasen et al., 2019; Becker-Weimann et al., 2013; Bischoff & Bryson, 1964). This uncoupling expands a cell's accessible behavioral repertoire beyond its canonical tissue identity, enabling the pursuit of local stress-minimizing strategies at the expense of collective form (Levin, 2019). As this process deepens, the first neoplastic cells become progressively dissociated from the electrical, mechanical, and chemical communication networks that sustain morphogenetic memory (Fig. 3A) (Becker-Weimann et al., 2013; Chernet & Levin, 2013a; Levin, 2012). The following sub-sections explore each channel of dissociation in turn, beginning with the loss of bioelectrical regulation.

2.2.1 Bioelectrical Dissociation

Cancer cells frequently downregulate or sequester connexins and other gap junction proteins, disrupting intercellular ionic exchange and synchrony (Garber et al., 1997; Mesnil et al., 1995; Ruch & Trosko, 2001; Yamasaki et al., 1995). This loss of coupling leads to depolarization of the resting membrane potential and desynchronization of calcium oscillations relative to the coordinated bioelectric state of the surrounding tissue (Chernet & Levin, 2013a, 2013b; Lobikin et al., 2012; Stewart et al., 2015). The resulting electrical isolation disconnects cancer cells from a key layer of the multicellular architecture that encodes and maintains morphogenetic memory (Mathews & Levin, 2017).

*2.2.2 Biomechanical Dissociation*

Loss of cadherin- (cell-cell) and integrin- (cell-matrix) mediated adhesion disrupts the transmission of mechanical forces through cytoskeletal networks, uncoupling cancer cells from both neighboring cells and the structural reference frame of the basement membrane (Chan, 2006; Hamidi & Ivaska, 2018; Ingber, 2002; Jeanes et al., 2008; Mizejewski, 1999). Coupled with the stress-induced suppression of mechanotransducive pathways, such as the Hippo cascade, this leads to a breakdown of the physical feedback loops that normally regulate growth and maintain tissue architecture (Misra & Irvine, 2018; Nelson & Bissell, 2006).

*2.2.3 Biochemical Dissociation*

Disruption of developmental morphogen gradients severs cancer cells from the positional information that maintains stable tissue identities (Polakis, 2012; Potter, 2007; Shang et al., 2017). Without these instructive chemical fields, cells dedifferentiate toward more plastic, stem-like states (Gawlik-Rzemieniewska & Bednarek, 2016; Kaufman et al., 2016; Novak et al., 2020) and frequently downregulate or mislocalize tissue-polarity proteins (Chan et al., 2006; Hatakeyama et al., 2014; VanderVorst et al., 2018). This biochemical uncoupling strips cells of both positional and directional identity, rendering them molecularly *disoriented* with respect to the surrounding tissue's regulatory architecture.

2.3 Breaking a cell

Once separated from the electrical, mechanical, and chemical cues that sustain multicellular memory, neoplastic cells regress into a unicellular interpretive structure dominated by genetic programs associated with immediate stress and survival (Bussey et al., 2017; Bussey & Davies, 2021; Zhou et al., 2018). This reprograming locks them into states of single-celled 'hyperarousal' (intrinsic stress generation) and 'hypervigilance' (extrinsic stress sensitivity)—consistently producing the hallmarks of cancer as downstream signaling consequences (Hanahan & Weinberg, 2000, 2011).

*2.3.1 Single-Cell Hyperarousal*

The hallmarks of cancer are actively sustained by "stress-addicted" circuitry: a core set of ancient adaptive programs, including reactive oxygen species (ROS)-sensitive signaling, hypoxia-inducible circuits, the unfolded protein response (UPR), and nutrient-stress-driven metabolic reprogramming (Fig. 3B) (Ummarino et al., 2024). Though these pathways evolved to help cells surmount transient environmental challenges, their persistent activation under conditions of unresolvable error traps neoplastic cells in a dissociated, oncogenic state (Tian et al., 2021).

Persistent ROS production activates key signaling nodes such as MAPK, JNK, NF-κB, and NRF2. Together, these pathways drive hallmark oncogenic traits: sustained proliferation, apoptotic resistance, replicative immortality, metabolic reprogramming, and genomic instability (Aggarwal et al., 2019; de la Vega et al., 2018). Concurrently, continuous stabilization of HIF-1α and HIF-2α promotes the expression of GLUT1, PDK1, LOX, Twist1 and Snail—driving energetic reprogramming, angiogenesis, metastatic niche formation, epithelial-mesenchymal transition, and invasion (Kao et al., 2016; Semenza, 2012; Wellmann et al., 2008). The PERK/eIF2α/ATF4 and IRE1α/XBP1 branches of the UPR, triggered by endoplasmic reticulum stress, further contribute to this survival state by promoting macromolecule autophagy, dampening antigen presentation, and reinforcing NRF2 activity to block cell death signaling (Granados et al., 2009; Hart et al., 2012; Hetz & Papa, 2018; Hsu et al., 2019).

These stress modalities converge on metabolic nodes associated with nutrient-stress, such as PI3K-Akt-mTOR, MYC, and AMPK, redirecting energy usage toward aerobic glycolysis, glutamine catabolism, and macropinocytosis (DeBerardinis & Chandel, 2016; Dengler, 2020; Gatenby & Gillies, 2004; Pavlova & Thompson, 2016). Though they may be surrounded by an adequately nourished organ, local conditions of nutrient scarcity cause cancer cells to exist in a persistent state of pseudo-starvation (García-Jiménez & Goding, 2019; Grasmann et al., 2019; White et al., 2015). Collectively, these chronic stress signals converge upon eIF2α and perpetuate the Integrated Stress Response (ISR), reprogramming cancer cells to exist in a near-constant state of survival at the level of single-cell physiology (Pakos-Zebrucka et al., 2016).

*2.3.2 Single-Cell Hypervigilance*

In addition to intrinsic stress generation, cancer cells exhibit enhanced sensitivity to paracrine and endocrine physiological error cues—signals which normally indicate homeostatic discrepancies to sustain body-wide tissue coordination. In this dissociated, "hypervigilant" state, these external sources of error come to reinforce the canonical hallmarks of cancer (Fig. 3B) (Hanahan & Weinberg, 2011).

Across many cancers, growth factor receptors (GFRs) such as EGFR, IGFR, FGFR, PDGFR and HGFR are upregulated or constitutively activated, hypersensitizing cells to morphological error markers. This enhanced signal detection feeds into PI3K/Akt, RAS/RAF-MAPK, and JAK/STAT pathways, sustaining decontextualized growth (Alfaro-Arnedo et al., 2022; Demoulin & Essaghir, 2014; Raj et al., 2022; Uribe et al., 2021; Xie et al., 2020). Simultaneously, Toll-like receptors (TLRs) are overexpressed in a wide variety of cancers, sensitizing cells to subtle signs of tissue damage (Droemann et al., 2005; Fukata et al., 2007; Goto et al., 2008; Helminen et al., 2014; Ochi et al., 2012; Schmaußer et al., 2005). By hyper-sensing damage-associated molecular patterns (DAMPs), such as HSPs, HMGB1, and fragmented ECM components, these receptors trigger inflammatory cascades that promote survival, immune evasion, and angiogenesis (Bohnhorst et al., 2006; Huang et al., 2005; Macedo et al., 2007; Pradere et al., 2014; Rakoff-Nahoum & Medzhitov, 2009).

Cancer cells also become hyper-attuned to stress-related neuroendocrine cues. β-adrenergic receptors (β-ARs)—which normally mediate the body's acute stress response—become upregulated in many tumors, sensitizing cells to catecholamines such as norepinephrine (Cole & Sood, 2012; Powe et al., 2011; Rains et al., 2017). β2-AR activation sustains proliferative signaling through ERK1/2, degrades p53 through β-arrestin-1/MDM2, and reinforces stemness through LDHA-dependent metabolic reprogramming (Cui et al., 2019; Flint & Bovbjerg, 2012; Hara et al., 2011; Wang et al., 2022). In addition, deregulated glucocorticoid receptor expression sustains immunosuppressive, anti-apoptotic states via NF-κB, Bcl-xL, and PD-L1 (Deng et al., 2021; Godbout & Glaser, 2006; Vilasco et al., 2011). In hormone-sensitive cancers, estrogen and androgen receptor hyperactivation amplifies responsiveness to developmental error cues via RAS/MAPK and PI3K/Akt pathways (Clusan et al., 2023; Heinlein

& Chang, 2004), while insulin and thyroid hormone receptor hyperactivation enhances sensitivity to energetic error cues through PI3K/Akt-mediated mitogenic and metabolic growth programs (Belfiore & Malaguarnera, 2011; Kim & Cheng, 2013). Together, these molecular alterations suggest a concerted reprogramming of cancer cells whereby hypervigilant extracellular sensory apparatuses come to amplify the detection of physiological stress signals, further reinforcing tumorigenic growth.

*2.3.3 Invasion*

This reframing of cancer suggests that invasion is driven by a fragmentation of 'perspective' between the tumor and surrounding tissue. Morphogenetically dissociated and driven by persistent stress signaling, cancer cells enact localized error-correction strategies—proliferation, ECM remodeling, and angiogenesis—which damage the surrounding tissue and enhance the generation of extracellular stress signals. While these cues normally lead to collective resolution, they come to be interpreted by the tumor as signs to accelerate proliferation and growth. As morphogenetic error-minimization tends towards infinity, tumors come to recapitulate the behavior of wounds that never heal (Dvorak, 1986). Caught in cycle of unresolvable error, the tumor is driven towards deeper dissociation, invasion, and secondary fragmentation in the form of metastasis.

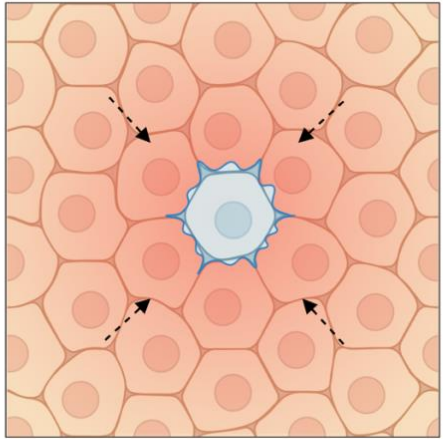
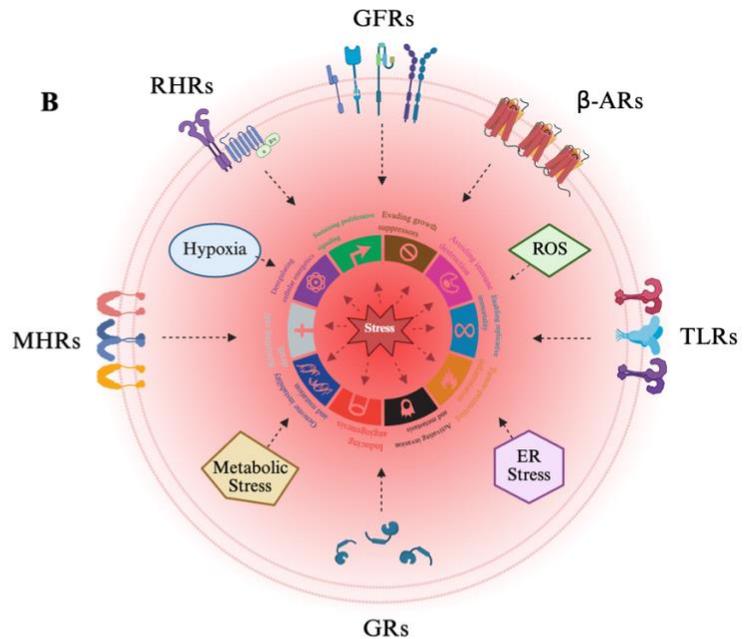
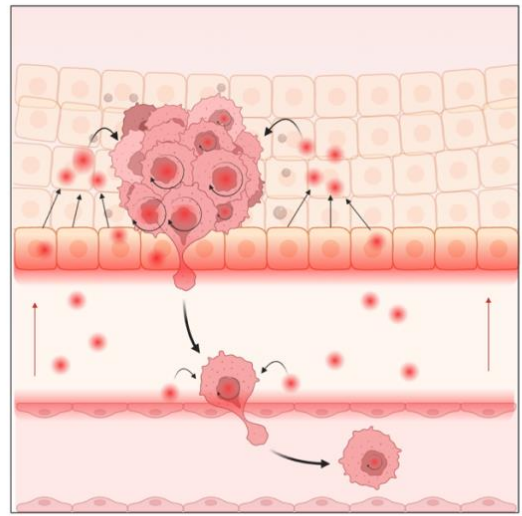

**Fig. 3: The Tumorigenic Loop**

**A,** Neoplastic Dissociation. After a critical threshold of stress, the first neoplastic cell will disconnect from the electrical, mechanical, and chemical networks that regulate tissue identity. Thereafter, the surrounding organ becomes functionally fragmented into distinct systems of morphogenetic memory. **B,** Hyperarousal and Hypervigilance. The hallmarks of cancer are collectively driven by sustained states of internal stress generation and external stress sensitivity. Reactive oxygen species (ROS), hypoxia, metabolic stress, and endoplasmic reticulum (ER) stress mediate internal "hyperarousal;" growth factor receptors (GFRs), β-adrenergic receptors (β-ARs), toll-like receptors (TLRs), glucocorticoid receptors (GRs), metabolic hormone receptors (MHRs), and reproductive hormone receptors (RHRs) mediate external "hypervigilance." **C,** Reinforcement and Invasion. Cancer cells damage the surrounding tissue, leading to the production of increased stress signals (red, hazy dots) which drive the tumor's underlying growth logic. In this infinite error-minimization loop, tumors come to resemble wounds that never heal.

*2.3.4 Transition to Psychology*

    The principle of stress-induced dissociation is not just confined to somatic tissues. It has been proposed that non-neural bioelectrical coordination provided the evolutionary substrate for neural signaling, scaling the architecture of collective regulation from tissues to minds (Levin, 2023). Across both levels, the integrity of shared memory networks is essential for aligning local perception with collective goals. Within this architecture, stress-induced dissociation appears to have been conserved across evolution: in simple Metazoan collectives, it enabled survival under unresolvable stress, while in complex neural ensembles it was exapted as a short-term adaptive strategy to enable the persistence of the entire organism. In this light, the dissociation that drives

tumorigenesis finds its cognitive analogue in traumatic dissociation, where neural circuits lose access to a coherent autobiographical reference frame. To develop this parallel, we now turn to the psychobiology of trauma.

## 3. Psychological Trauma: The Stress-Induced Dissociation of the Personality

Traumatic dissociation—often defined as the 'fragmentation of the Self'—occurs because we are all, at our core, collective intelligences (Fisher, 2017; Lanius et al., 2020; McMillen & Levin, 2024; Van der Kolk, 2014). One influential framework that explains this fragmentation is known as the Structural Dissociation of the Personality (Nijenhuis et al., 2010). According to this model, animals evolved psychobiological action systems—neural circuits that integrate perception, memory, and behavior—to coordinate adaptive responses such as attachment, defense, exploration, and energy regulation (Fanselow & Lester, 2013; Panksepp, 2004). Though each system selectively tunes sensory perception toward specific goals, they are normally integrated through shared memory networks spanning implicit, semantic, and episodic domains.

Under typical conditions, moderately stressful or surprising events lead to the formation of episodic memory—temporally structured representations with initiation and resolution (Roozendaal et al., 2009; Talarico et al., 2004). This process is orchestrated by coordinated activity across the thalamus (sensory integration), the amygdala (threat appraisal), and the hippocampus (contextual encoding and consolidation) (Aggleton & Brown, 1999; Richter-Levin & Akirav, 2000). The medial prefrontal cortex (mPFC) plays a key role in linking these discrete experiences across time to generate autobiographical memory, thereby constructing a coherent narrative sense of Self (Fig. 4A) (Damasio, 1999).

However, this normal process of memory formation break down under traumatic conditions (Fig. 4A)—defined as situations of extreme, unresolvable stress in which the organism's defensive strategies fail to resolve the danger (Keane et al., 2006; Nijenhuis et al., 1998). Such conditions may emerge acutely, as in physical violence, or chronically, as in prolonged relational violations (Lima et al., 2010; Stovall-McClough & Cloitre, 2006). Neurobiologically, trauma is associated with the hyperactivation of the amygdala, which disrupts top-down regulation from the mPFC and interferes with thalamic sensory gating and hippocampal encoding (Roozendaal et al., 2009; Shin et al., 2006). This leads to the fragmentation of memory: the event is not consolidated into coherent autobiographical stores but instead persists as discrete, sensorimotor traces—flashbacks, autonomic responses, and emotional reactivities–that resurface involuntarily and feel divorced from time (Fig. 4B) (Van der Kolk, 2014).

Over time, this fragmentation can result in structural dissociation, wherein action systems segregate into functionally distinct containers that process experience separately. As proposed by Nijenhuis, van der Hart, and Steele, this split commonly occurs between the action systems oriented toward defense and emotional attachment, which coalesce into the Emotional Personality (EP), and the systems responsible for daily adaptive functioning (e.g. social interaction, energy regulation), which form the Apparently Normal Personality (ANP) (Nijenhuis et al., 2010). The EP stores traumatic memory in an implicit, unprocessed form and remains hypervigilant to threat cues, while the ANP maintains outward functioning by dissociating from the traumatic material. Thus, the dynamics between the EP and ANP form a self-perpetuating loop of dissociation, arousal, and intrusion, in which fragmentary trauma memories periodically break through into awareness, overwhelming the ANP's regulatory capacity and reinforcing the split between systems (Fig. 4C) (Nijenhuis et al., 2010; Reinders et al., 2003).

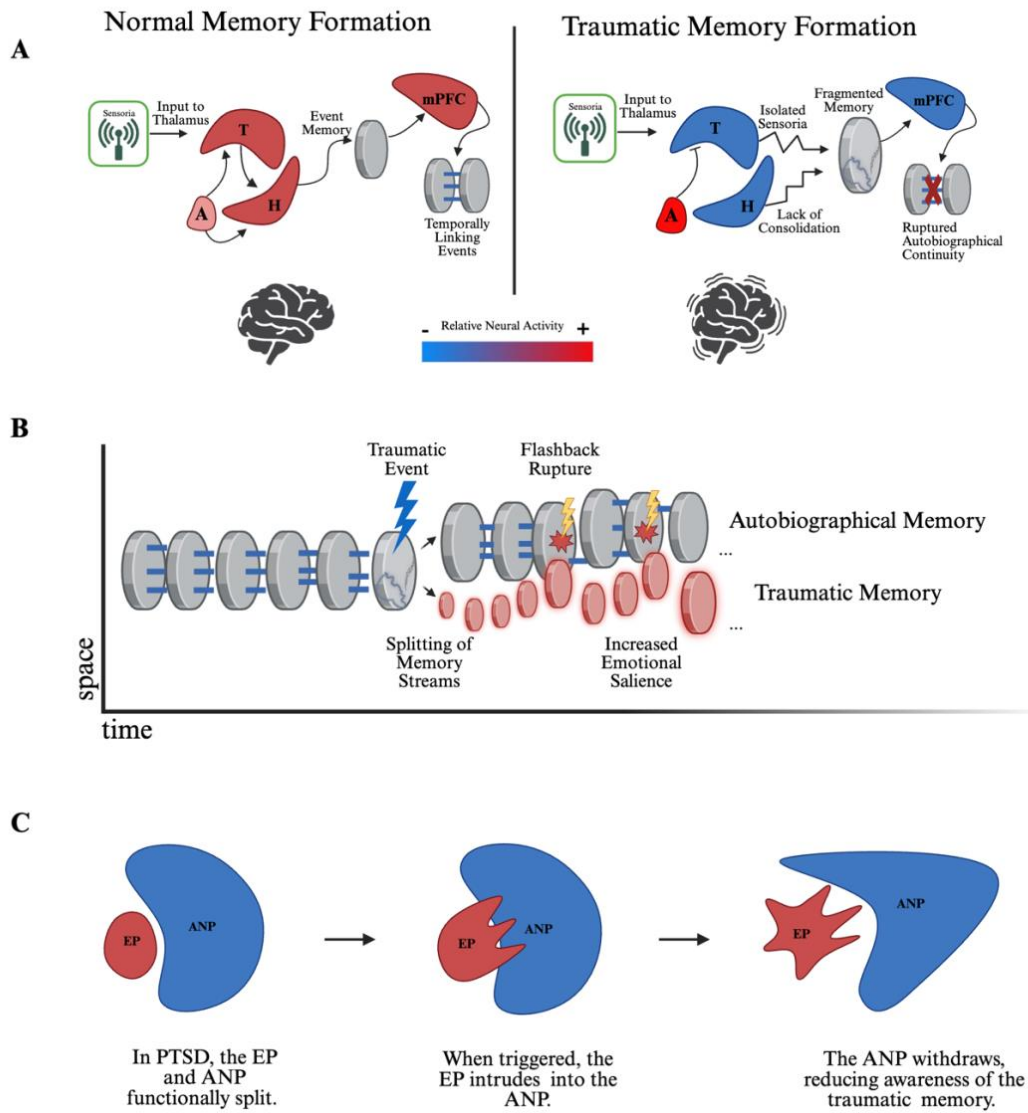

**Fig. 4: Dynamics of Traumatic Memory**

**A**, In normal settings, the amygdala(A)-hippocampal(H)-thalamic (T) axis integrates diverse sensory channels into cohesive event memories which are temporally linked by the medial prefrontal cortex (mPFC) to create a continuous sense of Self. During conditions of traumatic stress, amygdala hyperactivation breaks this circuit—fragmenting sensory experience into isolated channels and preventing the mPFC from contextualizing events as related to Self. **B**, after an overwhelming, unresolvable stress, fragments of traumatic memory become isolated in an adjacent memory system with respect to the rest of autobiographical storage. During traumatic flashback, these memories intrude into normal waking consciousness, temporally rupturing autobiographical continuity. **C**, dynamics of Emotional Personality (EP) vs. Apparently Normal Personality (ANP). Most of the time, the EP and ANP remain in functionally separated systems of memory. When triggered by trauma-related stimuli, the EP intrudes into the ANP, causing the ANP to further dissociate from the trauma. While this dissociation is geared to maintain outward functionality, it further deepens the divide and exacerbates the cycle.

3.1 Dissociation

In clinical cases, the ANP and EP exhibit access to mutually exclusive memory networks across episodic, semantic and procedural domains (Brewin, 2001; Markowitsch, 1999). The EP implicitly encodes the traumatic memory as a non-integrated sensorimotor and affective complex, unavailable for narrative recall or temporal contextualization (Van der Kolk & Van der Hart, 1991). In order to function in daily life, the ANP actively dissociates from the EP, employing threat-avoidant strategies such as amnesia, emotional numbing, and behavioral avoidance (Markowitsch et al., 2000; Van der Hart & Brom, 2000). Conversely, the EP cannot escape the trauma and increasingly dissociates from the body, resulting in symptoms such as psychoform and somatoform dissociation, depersonalization, and derealization (Choi et al., 2017; Nijenhuis et al., 1996; Shilony & Grossman, 1993). Functionally severed from the ANP, the EP lacks the cognitive context to situate its internal sensoria as stemming from the past, leading to disorienting flashbacks that are physiologically and emotionally experienced as if the traumatic event were re-occurring in the present moment (Kearney & Lanius, 2024).

3.2 Hyperarousal & Hypervigilance
When trauma-related stimuli (TRS) reawaken implicit memory traces, the EP becomes locked in a state of hyperarousal and hypervigilance—a sustained stress response shaped by both neuroendocrine activation and perceptual distortion (Yehuda et al., 2015). Hyperarousal is characterized by elevated catecholamines, low cortisol (with enhanced reactivity), and suppression of parasympathetic tone, prioritizing short-term survival at the cost of long-term regulation (Meyer et al., 2016; Southwick et al., 1999; Yehuda, 2002). This physiological profile impairs the person's capacity to return to homeostatic baseline, even after the external threat has subsided (Shalev et al., 1992; Weston, 2014). In parallel, hypervigilance emerges as a behavioral mode of anticipatory scanning and defensive over-interpretation, in which TRS are amplified, and neutral cues are increasingly misclassified as dangerous (Rauch et al., 2000; Shalev et al., 2000). The world is thus perceived as unpredictable, alien, and unsafe, limiting the ability to learn or internalize signs of safety (Murphy, 2023; Van der Kolk, 1994). Over time, this "stress-addicted" neural circuitry traps the mind in cycles of reliving the traumatic event (Van Der Kolk et al., 1985).

3.3 Intrusion
When the EP is triggered by referential stimuli, the ANP becomes transiently deactivated, and the individual is overtaken by the autonomic and affective intensity of the unprocessed traumatic memory. Because the ANP lacks episodic access to the trauma and cannot identify the triggering stimulus, these flashbacks are experienced as uncontrollable and inexplicable—producing a profound sense of helplessness and disorientation (Nijenhuis et al., 2010). The sheer terror of the re-experienced memory reinforces the ANP's defensive drive to suppress integration with the trauma-holding EP, deepening the dissociative split (Janet, 1904; Nijenhuis, 1994). At the same time, the unpredictability of these intrusive episodes generates anticipatory stress, which sensitizes the ANP to surrounding cues. Through associative learning, previously neutral internal sensations and environmental stimuli that precede or follow the intrusion become new TRS which the ANP learns to avoid in an effort to prevent future reactivation (Titchener, 2014). Over time, the conditional expansion of trauma-related stimuli can lead to widespread behavioral restriction, social withdrawal, and diminished cognitive flexibility—a degenerative process known as post-traumatic decline (Janet, 1909).

# 4. Implications: Tumorigenesis is a Trauma Response

Given that this core dynamic of stress, dissociation, and intrusion exists across scales of cognition, the concept of trauma should be generalized beyond the neural realm to reflect a critical vulnerability of the multicellular condition. This would allow us to formally define tumorigenesis as a morphogenetic trauma response: a stress-induced dissociative loop which reinforces anatomically intrusive behavior with respect to the adjacent tissue. Just as the EP becomes disconnected from the autobiographical memory of the ANP—which often exhibits avoidant behavior toward the trauma—cancer cells become disconnected from the morphogenetic memory of the surrounding tissue—which often exhibits immunological tolerance toward the tumor (Bonaventura et al., 2019). Paralleling how the EP implicitly encodes traumatic memories through affective and autonomic responses, cancer cells seem to implicitly encode the memory of unresolvable stress through their survival-driven signaling architectures. And just as trauma-related stimuli cause the EP to intrude into the ANP—potentially leading to psychosocial atrophy in posttraumatic decline—local physiological 'error cues' drive cancer cells to invade the normal adjacent tissue—potentially leading to physiological atrophy in cancer-associated cachexia (Ferrer et al., 2023). If trauma represents the fragmentation of the psychological Self, tumorigenesis represents the fragmentation of the morphogenetic Self—the collective locus of memory, interpretation and meaning-making in anatomical space (Kohn, 2013; Levin, 2019).

4.1 Cancer therapy must change to reflect a trauma-informed approach.

Once we shift our conceptual framing of cancer cells from *irreversibly mutated* to *molecularly traumatized*—cells whose interpretive architectures have been rewired by chronic, unresolvable stress—our therapeutic strategies will evolve as well. Rather than seeking the direct destruction of cancer cells, we should aim to restore their capacity to *belong to the body* again. Therapeutically, this would mean facilitating their reintegration within the surrounding tissue's regulatory architecture, such that local stress perception becomes re-mapped to collective resolution. Once reconnected to this morphogenetic map, cancer cells should spontaneously initiate apoptosis, anoikis, or context-appropriate differentiation, thus resolving the tumorigenic phenotype.

To create this next generation of cancer therapies, oncology will need to adopt the same principles that guide the healing of trauma in psychotherapy. In the treatment of PTSD, for example, the primary goal is to facilitate the *reintegration of the Self* through the reprocessing of fragmented memory (Lanius et al., 2015). This typically proceeds through a three-step process: (1) establishing safety and trust, (2) inducing cognitive plasticity and (3) reprocessing the intrusive memory to bring narrative resolution (Fisher, 2017; Schwartz, 2021; Shapiro, 2017). Once resolved, the traumatic event can be safely reintegrated into autobiographical memory, bringing an end to the cycles of intrusion and dissociation (Ecker et al., 2012).

As we re-approach cancer therapy, we must recognize that the tumor is a pathological expression of a functionally fragmented tissue which requires reintegration to fully resolve. Whereas psychological trauma is resolved by re-encoding narrative memory—a structure with temporal polarity—morphogenetic trauma may require re-encoding anatomical memory—a structure with spatial polarity. As a new generation of 'psycho-oncologists,' it will be our task to guide this reintegration in ways that prevent re-traumatization and refractory neoplastic dissociation.

What evidence suggests that such reintegration might be possible? A growing body of studies continues to demonstrate that even highly transformed cells can regain normal behavior when placed in corrective microenvironments (Costa et al., 2009; Hendrix et al., 2007; Illmensee & Mintz, 1976; Kasemeier-Kulesa et al., 2008; Postovit et al., 2008; Stoker et al., 1966). In more recent work, it was shown that modulating the bioelectrical topology of epithelial tissues can normalize tumor-like structures bearing canonical oncogenic mutations (e.g., KRAS, BRAF$^{V600E}$) (Chernet & Levin, 2013a, 2013b; Lobikin et al., 2012). This represents a form of morphogenetic memory reprocessing, where electrical correction alone is sufficient to initiate spontaneous reintegration. It remains to be tested whether combining molecular signals of safety with bioelectrical reprogramming could enhance therapeutic efficacy.

4.2 Cellular reference frames may be an important concept for molecular biology.

*4.2.1 Cachexia as classical conditioning.*
Drawing an analogy from posttraumatic decline, cancer-associated cachexia may reflect a maladaptive form of homeostatic attenuation from the perspective of the tissue in response to tumor-driven stress. As previously established, morphostatic signaling circuits—such as those involving GFs and DAMPs—drive tumor invasion. Diverse tissues have also exhibited capacities for experientially driven behavioral modulation (Chen et al., 2022; Ellenbroek et al., 2013; Natoli & Ostuni, 2019; Rosenbaum et al., 1982; Schemann et al., 2020). By associating the expression of their normal homeostatic circuits with persistent tumor-derived stress, tissues may adaptively suppress the mechanisms which enable physiological integration. Over time, this may lead to systemic atrophy in the form of cachexia—even when total tumor volume is low (Ferrer et al., 2023).

*4.2.2 Dual nature molecules depend on interpretive contexts.*
The "dual nature" of molecules such as TGF-β may not be an intrinsic property of the signals themselves, but rather a function of the interpretive context—or reference frame—through which they are interpreted. In an integrated cellular collective, TGF-β acts as a morphological error signal, guiding coordinated action towards whole-organ maintenance. In contrast, when a cell becomes dissociated, TGF-β may instead be interpreted as a cue to fulfill local modes of homeostatic error resolution through proliferation and EMT (Neel et al., 2012). This divergence suggests that the degree of morphogenetic integration or dissociation constitutes a critical variable for interpreting molecular behavior. To fully understand the function of signaling pathways in health and disease, we must begin to map not only molecular inputs and outputs—but also the shifting relational architectures that give those signals meaning.

**Conclusion**
In this Perspective, I have argued that tumorigenesis is an anatomical trauma response. This does not claim that tumors come from psychogenic origins, but rather that stress-induced dissociation is a core feature of multicellular architectures, from somatic tissues to neural ensembles. We began by elucidating a theoretical model whereby the sharing of stress signals, in combination with cellular adaption, led to the creation of distributed regulatory architectures—a shared molecular perspective which maps local stress-perception to collective homeostatic resolution. Then, we derived how conditions of unresolvable error expose a structural vulnerability within these architectures, leading to the fragmentation of morphogenetic memory and neoplastic dissociation. Because neoplastic cells are separated from a collective point of reference, sustained stress-signaling and sensitivity naturally produce the hallmarks of cancer—a concerted network of adaptations which seek to minimize homeostatic error from the single-cell perspective. This multi-agent fragmentation belongs to an evolutionary picture which not only encompasses simple Metazoan collectives, but has etiological implications for psychological disorders of extreme stress. It follows that the intrusive dynamics of PTSD share parallels with the invasive dynamics of tumorigenesis for structurally similar reasons: they are both disorders of stress and memory, spanning morphological to psychological domains. Together, these arguments reframe tumorigenesis as a structurally dissociated, survival-driven, anatomically intrusive behavior—in short, a morphogenetic trauma response. Thus, to win the "war on cancer," oncologists must shift strategies to prioritize reintegration, reprocessing and resolution over direct tumor destruction.

**Acknowledgements**
I would like to thank Michael Levin for his extensive feedback throughout the process of writing this manuscript, as well as Alexey Tolchinsky for his insightful and constructive comments.
All figures were Created with Biorender.com.